\documentclass[journal]{IEEEtran}

\ifCLASSINFOpdf
\else
\fi
\usepackage[cmex10]{amsmath}
\usepackage{multirow}

\usepackage{url}

\ifCLASSOPTIONcompsoc
\usepackage[caption=false,font=normalsize,labelfont=sf,textfont=sf]{subfig}
\else
\usepackage[caption=false,font=footnotesize]{subfig}
\fi

\usepackage{cite}
\usepackage{color}
\usepackage{booktabs,colortbl}
\usepackage{amssymb}
\usepackage{setspace}
\usepackage{algorithm}
\usepackage[noend]{algpseudocode}

\ifCLASSINFOpdf
\usepackage[pdftex]{graphicx}
\graphicspath{{./Images/}}
\DeclareGraphicsExtensions{.pdf,.jpeg,.png}
\else
\fi

\begin{document}
%
\title{Optimization-based Settingless Algorithm Combining Protection and Fault Identification}


\author{Nadezhda Davydova, \IEEEmembership{Student Member, IEEE,
	} Dmitry Shchetinin, \IEEEmembership{Member, IEEE,} and Gabriela Hug,  \IEEEmembership{Senior Member, IEEE}}

\maketitle

\begin{abstract}
	The demand for faster protection algorithms is growing due to the increasingly faster dynamics in the system. The majority of existing algorithms require empirically selected set-points, which may reduce sensitivity to internal faults and cause security problems. This paper addresses these challenges by proposing a settingless time-domain unit protection algorithm for medium-voltage lines. The main idea of the algorithm is to identify which model of a protected line, i.e. healthy or with an internal fault, is more consistent with the input measurements. This is done by solving a number of small-scale convex optimization problems, which at the same time determine the characteristics of an internal fault that best fit the measurements. Thus, the proposed algorithm merges protection, fault location and fault type identification functionalities. The algorithm's performance is extensively tested on a grid model in MATLAB Simulink for different types of generation and grid operating conditions. The results demonstrate that the algorithm can operate quickly and  reliably, and accurately estimate fault characteristics even in the presence of noisy measurements and uncertain line parameters.
\end{abstract}


\section{Introduction} \label{INTRO} 
Deployment of distributed generation (DGs) and increasing power demand are pushing power systems to their limits and, as a consequence, are tightening the requirements for the fault clearing time \cite{ohrstrom_evaluation_2005}. In addition, converter-based DGs produce voltages and currents in case of a fault that may cause malfunctioning of conventional phasor-based relays. These problems necessitate the development and utilization of time-domain protection algorithms, which by their nature should operate well below one cycle of the power frequency and work reliably in grids with various types of DGs. Two main approaches have been proposed in the literature in this direction: traveling wave (TW) protection algorithms and protection algorithms based on differential equations for lumped circuits \cite{schweitzer_speed_2015}. \par 

The first approach is based on detecting and analyzing traveling waves in a power grid with line models represented by circuits with distributed parameters. Since fault-generated TWs propagate close to the speed of light and contain valuable information about their source, TW protection algorithms can be very fast and sensitive to grid disturbances. Various methods have been proposed in the literature to detect a faulty line using TW theory. For instance, the authors in \cite{costa_two-terminal_2017} propose a two-terminal TW protection that is based on the analysis of the arrival times of the first wavefronts of the disturbance-generated TWs. A differential protection based on equivalent TWs is presented in \cite{tang_new_2017}. An extensive review of these and other TW protection algorithms can be found in \cite{schweitzer_speed_2015} and \cite{schweitzer_time_2017}, in which their advantages and disadvantages are highlighted. A common problem for a majority of the TW protection algorithms is that they may malfunction in case of close-in faults and faults with small inception angles. \par

The second approach is based on analyzing differential equations that describe a lumped-parameter model of a power grid. While it is naturally slower than the TW-based approach, it does not have the aforementioned problems of the TW protection algorithms. The authors in \cite{lei_ultra-high-speed_2018,hashemi_transmission-line_2013,kong_study_2015} propose protection algorithms based on incremental quantities, which are fault-generated voltages and currents. Particular combinations of the incremental quantities are compared to thresholds to classify internal and external faults. A summary of algorithms based on incremental quantities is presented in \cite{schweitzer_speed_2015}. Taking it a step further, the authors in \cite{schweitzer_TD_TW_2017} propose a protection scheme that merges incremental quantity- and TW-based algorithms to improve the performance of time-domain protection. In \cite{schweitzer_setting_2017}, detailed guidelines for the calculation of set-points for this scheme are presented. Another protection principle that is based on dynamic state estimation is proposed in \cite{meliopoulos_setting-less_2013} and \cite{liu_dynamic_2017}. It utilizes measurements of terminal voltages and currents to estimate the goodness of fit of a healthy line model to the measurements. The goodness of fit obtained with the chi-square test is compared with a threshold to identify whether the protected line is healthy or not. While the proposed algorithm operates reliably in case of internal faults, transients due to certain external faults may lead to an erroneous tripping of a protected line. Therefore, a half-a-cycle delay is introduced before making a final tripping decision. \par

To the best of the authors' knowledge, all proposed and conventional protection algorithms require set-points. The majority of these set-points are selected empirically, which may lead to reduced sensitivity to internal faults and security problems. The aim of this paper is to address this drawback by proposing an optimization-based settingless unit protection algorithm for medium-voltage power lines. This time-domain algorithm belongs to the group of approaches based on differential equations for lumped circuits. The main idea of the algorithm is to identify which model of a protected line, healthy or with an internal fault, is more consistent with the input measurements. This is done by solving a number of optimization problems, which at the same time determine the characteristics of an internal fault that best fit the measurements. The proposed algorithm has the following distinctive features and advantages over the existing algorithms: 

\begin{itemize}
	\item relies on instantaneous voltage and current measurements from both line ends, 
	\item does not require selection of set-points, only parameters of the protected line have to be predefined,
	\item merges protection, fault location and fault type identification functionalities to unify these closely related functions and determine all key fault characteristics at once,
	\item is efficient and reliable since the optimization problems are small-scale and convex, hence they can be quickly solved to global optimality.
\end{itemize}


\section{Models of Protected Line} \label{Line_mod} 
\begin{figure}[!t] 
	\centering
	\includegraphics  [width=0.37\textwidth]{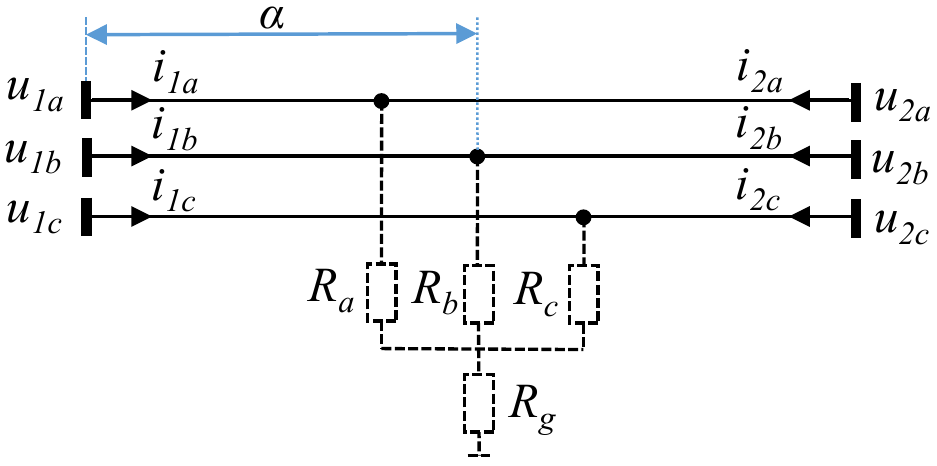}
	\vspace{-0.1in}
	\caption{Model of a power line with/without an internal fault}
	\label{fig:line_model}
	\vspace{-0.2in}
\end{figure}

This work assumes that the protected line can be in two states: healthy or with an internal fault, which is represented by a generic model shown in Fig.~\ref{fig:line_model}. Both line ends are connected to the grid, a representation of which is omitted since the proposed algorithm does not require any grid information. A healthy line at any time instance can be described with the following differential equations in matrix form:
\vspace{-0.05in}
\begin{align}  
&u_{1} - u_{2} - \left(R \cdot i_{1} + L \cdot \frac{\partial i_{1}}{\partial t}\right) =0 \label{eqn:1n} \\
&u_{2} - u_{1} - \left(R \cdot i_{2} + L \cdot \frac{\partial i_{2}}{\partial t}\right) =0, \label{eqn:2n} 
\end{align} \vspace{-0.07in}

\noindent where $R$ and $L$ are matrices of self and mutual resistances and inductances of phases, respectively, and $u_1$, $u_2$, $i_1$ and $i_2$ are vectors of three-phase instantaneous voltages and currents. Applying Kirchhoff's laws to the line model with an internal fault yields the following equations:
\vspace{-0.05in}
\begin{align}  
&u_{1} - u_{2} - \alpha \cdot Z \cdot i_{1} + (1-\alpha) \cdot Z \cdot i_{2} =0 \label{eqn:1f} \\
&u_{1} - \alpha \cdot Z \cdot i_{1} - Z_F \cdot (i_{1} + i_{2}) =0, \label{eqn:2f} 
\end{align} \vspace{-0.15in}   

\noindent where $Z := R + L \cdot \frac{\partial }{\partial t}$ and $\alpha$ is the fault location defined as relative distance from the left line terminal to the fault with respect to the line length. The matrix of fault resistances $Z_F$ is defined as follows: 
\begin{equation}
Z_F :=
\begin{bmatrix}
R_a+R_g & R_g & R_g \\
R_g & R_b+R_g & R_g \\
R_g & R_g & R_c+R_g 
\end{bmatrix}.
\end{equation}
Equations \eqref{eqn:1n}-\eqref{eqn:2f} serve as the basis for the proposed algorithm.  

\section{Proposed Algorithm} \label{MAIN_ALG}  

\subsection{Algorithm Overview} \label{MAIN_over}  
\begin{figure}[!t] 
	\centering
	\includegraphics  [width=0.38\textwidth]{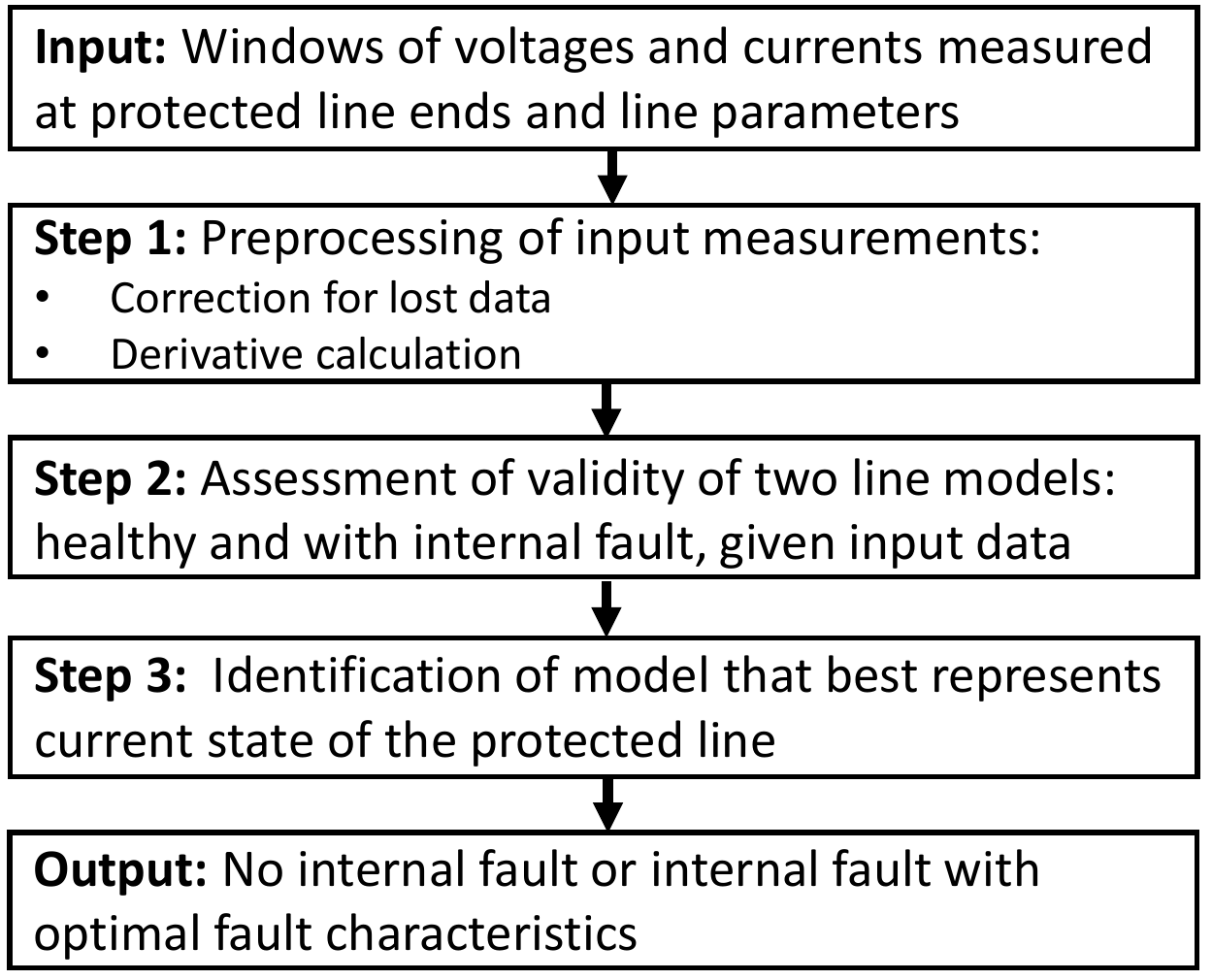}
	\vspace{-0.1in}
	\caption{Outline of proposed algorithm}
	\label{fig:outline}
	\vspace{-0.2in}
\end{figure}

The proposed algorithm is shown in Fig.~\ref{fig:outline} and is intended to perform settingless unit protection of a medium-voltage line and identification of fault characteristics. Its inputs are:
\begin{itemize}
	\item the parameters of the protected line (matrices $R$ and $L$ described in Section~\ref{Line_mod}), which are assumed to be accurately estimated using data-driven methods proposed in literature \cite{costa_estimation_2015,liao_online_2009};   
	\item short observation windows containing samples of instantaneous three-phase voltages and currents from both ends of the line, which must be synchronized.
\end{itemize}

The algorithm consists of three main steps. The first step is to preprocess the input measurements to account for the missing data and obtain time derivatives of measured currents, which the algorithm requires since it relies on differential equations of a power line. At the second step, the algorithm evaluates how well two models of a power line, i.e. healthy and with an internal fault, fit the input measurements. To do this, a number of optimization problems are solved, which allows to reliably detect an internal fault and identify its characteristics as the ones that are the most consistent with the data. Based on the evaluation carried out at the second step, the last step chooses the model that best represents the current state of the protected line. This results in the identification of whether the line is faulty or not and, if yes, the most probable fault characteristics given the data. A detailed description of these steps is given below along with the hardware requirements for the implementation of the proposed algorithm. 

\subsection{Preprocessing of Input Measurements} \label{MAIN_der} 
The proposed algorithm utilizes instantaneous measurements of voltages and currents and their derivatives for evaluation of healthy and faulty line models, which are based on differential equations. To obtain the correct evaluation, the windows of measurements have to be time aligned and their derivatives reliably estimated. This is ensured by the preprocessing step of the proposed algorithm. \par

At this step, the proposed algorithm initially checks if any samples of measurements from the other end of the line are lost during the data exchange via communication channels. If missing data is detected, the samples that correspond to the time stamps of this data are excluded from all the obtained windows of measurements. In this way the proposed algorithm ensures that all samples of measurements are time aligned. Note that by design of the second step of the algorithm, this procedure has no negative impact on the performance of the protection and fault identification functions provided that a small percentage of samples are lost. The methods for identification of lost data packages are covered in \cite{ff} and therefore, outside of the scope of this paper.   \par

Next, the proposed algorithm determines the time derivatives of measured currents. This can be done by fitting successive sub-sets of a predefined number $l\ge3$ of adjacent samples with a second-degree polynomial using linear least squares. The obtained polynomials are then differentiated at the time instances corresponding to the time stamps of the measurements' samples. This method filters out high frequency noise while not impairing the quality of the derivative estimation. \par

\begin{figure}[!t]  
	\vspace{-0.18in}
	\centering
	\subfloat[Inception moment at $t=100$]{\includegraphics[width=1.65in]{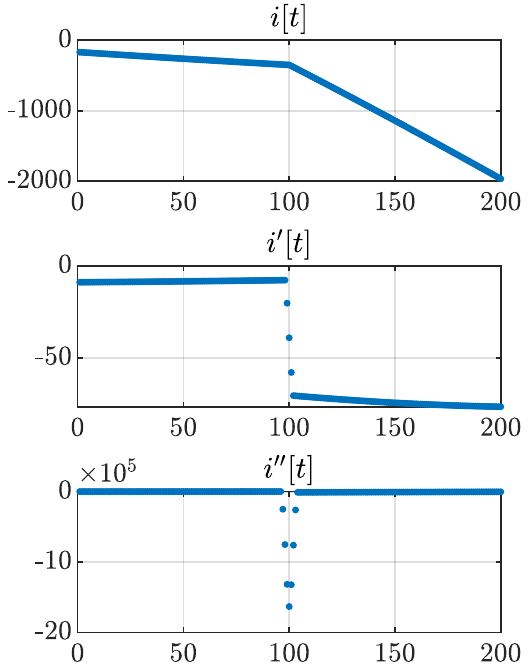}
		\label{fig_I_and_dI_fault}}
	\hfil
	\subfloat[No inception moment]{\includegraphics[width=1.62in]{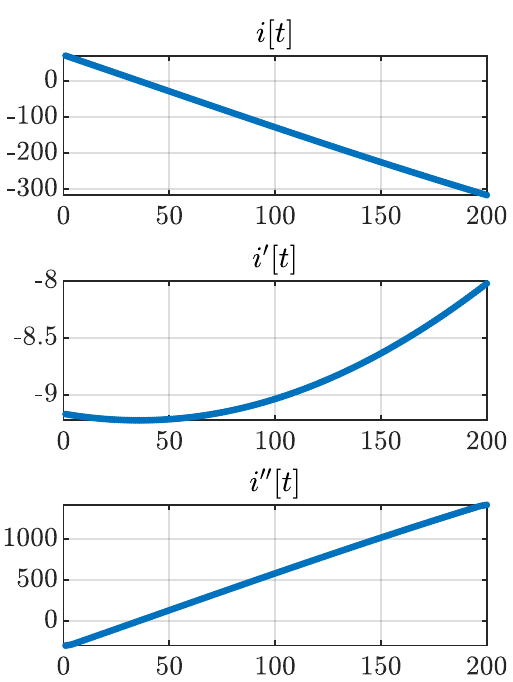}
		\label{fig_I_and_dI_no_fault}}
	\caption{Measured current and its numerically estimated derivatives for $l=5$}
	\label{fig_I_and_dI}
	\vspace{-0.2in}
\end{figure}

Unfortunately, this method may incorrectly calculate the derivatives for several samples around the fault inception moment. To see this, consider a window of samples of the measured phase current with the fault occurring in the middle of the window, shown in Fig.~\ref{fig_I_and_dI_fault}. At the fault inception moment, the true derivative of the current is undefined due to a sudden change of the line model. While the numerical estimate of this derivative always exists, it is incorrect whenever it is computed using the samples both before and after the inception moment. Therefore, there are $l-2$ samples for which the derivative estimate is erroneous. To enable reliable operation of the algorithm, they must be identified and removed from the windows of measurements and derivatives. \par

The identification of such samples is based on the numeric estimate of the second derivative of the current. Since it shows the rate of change of the first derivative, it has the maximum absolute value for the samples closest to the fault inception moment (see Fig.~\ref{fig_I_and_dI_fault}). Hence, $l-2$ consecutive samples with the highest absolute values of the numerically estimated second derivative are removed from the observation windows. Note that these samples are removed even if the observation windows do not contain the fault inception moment as in Fig.~\ref{fig_I_and_dI_no_fault}. However, it has virtually no negative impact on the protection and fault identification functions in the case with no fault provided that $l$ is small. It is worth noting that the wavelet transform can be alternatively used for detection of erroneous estimates of derivatives since it is well-known for finding sudden changes in the signals \cite{kim_wavelet_2000}. However, an empirical analysis indicated that the proposed method based on the second derivative has better performance than the wavelet transform for this task. \par

This step of the algorithm results in matrices $U_1$, $U_2$, $I_1$, $I_2$, $I'_1$, $I'_2\in\mathbb{R}^{3\times N}$ of samples of measured voltages and currents from both line terminals as well as the computed estimates of current derivatives. Here, $N$ denotes the total number of timestamps that remain after accounting for the missing data and errors in the derivative estimation. Each row of the matrices corresponds to a particular phase and each column corresponds to a particular time stamp. The obtained matrices are further utilized at the second step of the algorithm.  

\subsection{Evaluation of Line Models} \label{MAIN_line_mod} 
\begin{figure}[!t] 
	\centering
	\includegraphics  [width=0.43\textwidth]{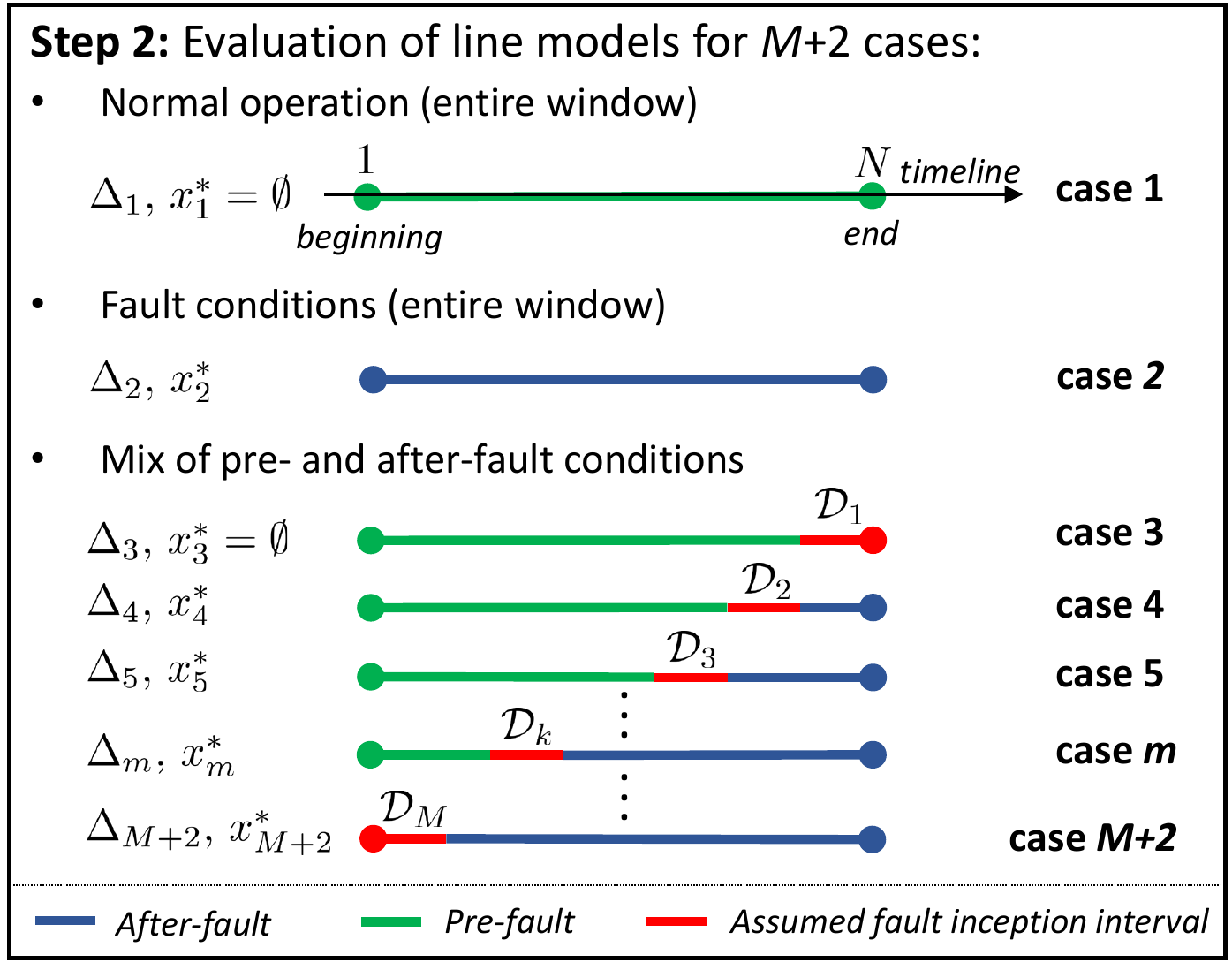}
	\vspace{-0.03in}
	\caption{Outline of second step of proposed algorithm}
	\label{fig:step2}
	\vspace{-0.2in}
\end{figure}

This is the main step of the algorithm as it is used to quantitatively evaluate how well the measurements fit two models: a healthy line and a line with an internal fault, described by equations \eqref{eqn:1n}-\eqref{eqn:2n} and \eqref{eqn:1f}-\eqref{eqn:2f}, respectively. For the latter model this step also identifies the internal fault characteristics that best correspond to the input data. The proposed algorithm determines the fault location, type, values of fault resistances and inception moment. The main idea behind this step is to consider $(M+2)$ possible hypotheses that reflect the potential nature of the measurements in the observation window as shown in Fig.~\ref{fig:step2}:
\begin{itemize}
	\item all samples correspond to the normal operation of a protected line,
	\item all samples correspond to the time period after the inception of an internal fault, 
	\item $M \geq 2$ possible cases of mixtures of samples before and after an internal fault in the observation window. In principle, $M$ should be equal to $N$ since an internal fault can occur at any moment in the window. However, to limit the number of considered cases, it is assumed that at each mixture case a fault has occurred within a particular interval of $R$ timestamps\footnote{Since $M$ must be a natural number, the values of $N$ and $R$ have to be selected accordingly.}, which thus gives $M=\frac{N}{R}$.
\end{itemize}
Clearly, one of these $M+2$ hypotheses must be true. Finding it requires quantitative assessment of the validity of all hypotheses. The validity of the $m$-th hypothesis is quantified through a mean squared error, denoted by $\Delta_m$, of the corresponding line model (healthy or faulty) or their mixture given the set of input measurements. Hence, the proposed algorithm constructs matrices $S$ and  $W(R_F,\alpha)\in\mathbb{R}^{6\times N}$ of mismatches of equations  \eqref{eqn:1n}-\eqref{eqn:2n} and \eqref{eqn:1f}-\eqref{eqn:2f}, respectively, defined as follows:
\begin{equation}  \label{S} 
S:=\begin{bmatrix}
\,U_{1} - U_{2} - \left(R \!\cdot\! I_{1} + L \!\cdot\! I'_{1}\right)\, \\[0.3em] \,U_{2} - U_{1} - \left(R \!\cdot\! I_{2} + L \!\cdot\! I'_{2}\right)\,
\end{bmatrix} 
\end{equation}
\begin{equation} 
W(R_F,\alpha):=\begin{bmatrix} \label{W}
\,U_{1} - U_{2} - \alpha\! \cdot\! Z\! \cdot\! I_{1} + (1-\alpha)\! \cdot\! Z \!\cdot\! I_{2}\, \\[0.3em] U_{1} - \alpha \!\cdot\! Z \cdot I_{1} - Z_F \!\cdot\! (I_{1} + I_{2})
\end{bmatrix}, 
\end{equation}
where $R_F \, := \, \left[\, R_a, \, R_b, \, R_c, \, R_g \,\right]^T$ is the vector of fault resistances shown in Fig~\ref{fig:line_model}. Note that while $S$ is a constant matrix because its entries are purely based on the measurement values, the elements of $W$ depend on the values of $R_F$ and $\alpha$, which are unknown. The procedure for computing $\Delta_m$ for all considered cases and estimating $R_F$ and $\alpha$ that fit the measurements the best is described below.

 
\subsubsection{Normal operation (case 1 in Fig.~\ref{fig:step2})} \label{Line_norm}  First, the proposed algorithm assesses the case that assumes that all samples in the observation window correspond to a healthy line. Hence, the value of $\Delta_{1}$ is computed as follows:
\begin{equation} \label{eqn:Normal_mis}  
\Delta_{1} := \frac{1}{6N} \sum_{i = 1}^6 \sum_{j=1}^N S_{i,j}^2,
\end{equation} 
The value of $\Delta_{1}$ serves as a measure of validity of this hypothesis given the input data. \par

\subsubsection{After-fault operation (case 2 in Fig.~\ref{fig:step2})} \label{Line_fault} Next, the case that assumes that all samples in the observation window correspond to the time period after the internal fault inception is considered. For this case, the input data is fitted to the line model with a fault, described by \eqref{eqn:1f}-\eqref{eqn:2f}. Since the fault characteristics are unknown, the proposed algorithm computes the mean squared error $\Delta_2$ of the faulty line model by determining the values of $R_F$ and $\alpha$ that minimize $\Delta_2$. This is done by solving the following optimization problem: \vspace{-0.065in} 
\begin{subequations}    \label{eq:opt_fault}
	\begin{alignat}{2}
	\Delta_2:=\;& \underset{R_F, \alpha}{\text{minimize}} && \enspace \frac{1}{6N}\sum_{i = 1}^6 \sum_{j=1}^N W_{i,j}^2(R_F,\alpha)  \label{eq:opt_fault_objective} \\[-0.1em]
	& \text{subject to} &&\enspace \enspace 0\le R_F\le R_F^{\max} \label{eq:opt_fault_con1}\\[-0.1em]
	& &&\quad 0\le\alpha\le 1,  \label{eq:opt_fault_con2}
	\end{alignat}
\end{subequations} 
where $R_F^{\max}$ is an upper bound placed on $R_F$ in order to consider only realistic values of fault resistances and improve numerical stability of the solution algorithm. Problem \eqref{eq:opt_fault} can be rewritten in a standard form as \par \vspace{-0.065in} 
\begin{subequations}  \vspace{-0.1in}  \label{eq:fault} 
	\begin{alignat}{2} 
	\Delta_2:=\;& \underset{x}{\text{minimize}} && \enspace  x^T \cdot H \cdot x + F^T\cdot x + d \label{eq:min_ext_objective_k}  \\ 
	& \text{subject to} &&  \enspace 0 \le x \le x^{\max},  \label{eq:fault_bound} 
	\end{alignat}  
\end{subequations} 
where $x:=\left[\, R_F, \,\alpha \,\right]$, $x^{\max}:=\left[\, R_F^{\max}, \,1 \,\right]$, $H\in\mathbb{R}^{5\times5}$, $F\in\mathbb{R}^5$, and $d\in\mathbb{R}$.
It follows from \eqref{W} and \eqref{eq:opt_fault_objective} that $H$ is a positive-definite matrix, hence problem \eqref{eq:fault} is convex and can be solved reliably to global optimality by any commercial solver. The obtained value of $\Delta_{2}$ represents the smallest possible mean squared error of the faulty line model. Note that the optimal value of the variable vector, denoted by $x^*_2$, provides the fault characteristics that best fit the input data. \par 

\subsubsection{Mixed operating conditions (cases 3 to M+2 in Fig.~\ref{fig:step2})} \label{Line_mix} 
Next, $M$ cases are assessed in which it is assumed that the observation window includes measurements from before and after an internal fault. Let $1\le k\le M$ be the index of a given mixture case, thus $k=m-2$. Further, let $\mathcal{P}$ be the set of the indices of all timestamps in the observation window, i.e. $\mathcal{P}:=\left\lbrace\, 1,\, 2,\, \dots, \, N \, \right\rbrace $. As mentioned above, the proposed algorithm assumes that for each mixture case $k$ the fault has occurred within a particular time interval containing $R$ timestamps. Set $\mathcal{P}$ can be then divided into the following three disjoint sets:
\begin{itemize}
	\item set $\mathcal{N}_k$ of the indices of timestamps corresponding to the assumed time interval before the fault;
	\item set $\mathcal{D}_k$ of the indices of timestamps corresponding to the assumed time interval of the fault inception;
	\item set $\mathcal{F}_k$ of the indices of timestamps corresponding to the assumed time interval after the fault.
\end{itemize}
This idea is illustrated in Fig.~\ref{fig:step2}, in which set $\mathcal{N}_k$ corresponds to the green part of the observation window, set $\mathcal{D}_k$ to the red part, and set $\mathcal{F}_k$ to the blue part. These sets are defined for the $k$-th mixture case as follows:
\begin{align} 
\mathcal{N}_k &:=  \begin{cases}
\emptyset, &  \text{if } k=M \\
\left\lbrace j \in  \mathbb{N} \, \left\lvert \!
\begin{array}{l}
j\ge 1 \\ 
j\le N (\frac{M-k}{M})
\end{array}
\right. \!\!\! \right\rbrace, &  \text{otherwise}  
\end{cases} \label{eq:Nm} \\
\mathcal{D}_k &:= \left\lbrace j \in  \mathbb{N} \, \left\lvert \!
\begin{array}{l}
j\ge N (\frac{M-k}{M}) +1 \\ [0.35em]
j\le N (\frac{M-k+1}{M})
\end{array}
\right. \!\!\! \right\rbrace  \label{eq:Dm} \\
\mathcal{F}_k &:=  \begin{cases} 
\emptyset, &  \text{if } k=1 \\
\!\left\lbrace j\! \in  \mathbb{N} \, \left\lvert \!
\begin{array}{l}
j\ge N (\frac{M-k+1}{M})+1\\ 
j\le N
\end{array}
\right. \!\!\! \right\rbrace\!, \!\!\!\!&  \text{otherwise.} 
\end{cases} \label{eq:Fm}
\end{align}
Naturally, it holds that $\mathcal{N}_k \cup \mathcal{D}_k \cup \mathcal{F}_k=\mathcal{P}$ for $\forall k$. Note that by design of \eqref{eq:Dm} sets $D_k$ are selected in such a way that $\,\bigcup_{k=1}^{M} \mathcal{D}_k = \mathcal{P}$ and $\mathcal{D}_i\cap\mathcal{D}_j=\emptyset$ for all $i\neq j$. This ensures that for every possible fault inception moment in the observation window there exists one corresponding mixture case whose assumed fault inception interval contains this moment. \par

The mean squared error of each mixture case is computed by fitting the green part of the measurement windows to the healthy line model and the blue part to the faulty line model. The red part of each observation window is not utilized in fitting since there is an uncertainty which of the two line models this part corresponds to. Therefore, the following optimization problem is solved to obtain the value of $\Delta_m$: \vspace{-0.2in}
\begin{subequations}    \label{eq:opt_mixed}
	\begin{alignat}{2}
	\!\!\Delta_m:=\;& \underset{R_F, \alpha}{\text{minimize}} && \enspace \frac{1}{\eta}\sum_{i = 1}^6 \!\left(\!\!\sum_{\,\,\,j\in \mathcal{N}_k}\!\!\!S_{i,j}^2+\!\!\!\sum_{j\in \mathcal{F}_k}\!\!W_{i,j}^2(R_F,\alpha)\!\right) \! \label{eq:opt_mixed_objective} \\[-0.1em]
	& \text{subject to} &&\enspace \enspace 0\le R_F\le R_F^{\max} \label{eq:opt_mixed_con1}\\[-0.1em]
	& &&\quad 0\le\alpha\le 1,  \label{eq:opt_mixed_con2}
	\end{alignat}
\end{subequations}
where $\eta=6(N-R)$. Similarly to \eqref{eq:opt_fault}, problem \eqref{eq:opt_mixed} can be rewritten in a standard form given by \eqref{eq:fault} and is also convex. Therefore, it can be solved fast and reliably to obtain $\Delta_m$, which is a measure of validity of the $k$-th mixture case (or $m$-th case among all $M+2$ cases). The optimal value of the variable vector $x^*_m:=[R_F^*,\alpha^*]$ provides the fault characteristics that best fit the input data. \par

Note, the values of $N$ (length of the observation window) and $M$ (number of considered mixture cases) must be provided by the user. These values control the trade-off between the speed of the algorithm and the accuracy of estimating the fault characteristics. A smaller $N$ reduces the amount of data that has to be stored and transmitted over the communication channel, while a smaller $M$ decreases the number of mixture cases to be evaluated. On the other hand, higher values for $N$ and $M$ improve the accuracy of estimating $R_F$ and $\alpha$ at the expense of increasing the computational burden. The selection of optimal values for $N$ and $M$ is outside of the scope of this paper as it depends also on the capabilities of the protection devices and the communication channel.

\subsection{Comparison of Evaluated Cases} \label{MAIN_last} 
Once all $M+2$ cases have been processed, the algorithm proceeds to the selection of the case that best corresponds to the data in the measurement windows. This case is identified as the one that has the smallest mean squared error $\Delta_m$. If the model with no internal fault is determined to be the most likely one, the algorithm takes new measurement windows and repeats the procedure described above. Otherwise, it returns an estimate of the fault characteristics and sends a tripping signal to the corresponding circuit breaker. Hence, the proposed algorithm simultaneously performs protection, fault location and fault type detection for a power line. All these functions are crucial for stable and reliable operation of the grid. For instance, fast and accurate fault location performed by the proposed algorithm can reduce the grid restoration time. Fault type detection accomplished by identifying $R_F$ is important for power lines with single pole switching \cite{mm}. In addition, the values of $R_F$ can be utilized for collecting statistical records on faults. \par

It is important to note that for both cases $m=1$ and $m=3$, the faulty line model is not utilized, i.e. the mean squared errors are computed using only the healthy line model (see Fig.~\ref{fig:step2}). The difference between these two cases is that all $N$ timestamps are used for computing $\Delta_1$, whereas $\Delta_3$ is obtained using only $N-R$ timestamps. Hence, limited computer precision and the presence of noisy measurements could result in $\Delta_3<\Delta_1$ even though in reality no internal fault has occurred within the observation window. To enable reliable operation of the proposed algorithm in such situations, an extra step is carried out whenever $\Delta_3$ and $\Delta_1$ have the smallest and the second smallest values among all $\Delta_m$. The idea is to compare the relative validity of case $m=1$ with respect to cases $m=3$ and $m=4$ in order to identify whether case $m=1$ is closer to $m=3$ (candidate for the true hypothesis) or case $m=4$ (false hypothesis). To do this, parameters $a_1:= \Delta_1/\Delta_3$ and $a_2:= \Delta_4/\Delta_1$ are computed and compared to each other. If $a_1 < a_2$, the algorithm concludes that the line is healthy. Otherwise, the occurrence of an internal fault at the time interval corresponding to $\mathcal{D}_1$ is selected as the valid hypothesis.

A similar procedure is utilized for cases $m=2$ and $m=M+2$, since in both of these cases the mean squared error is computed using only the faulty line model.

\subsection{Hardware Requirements for Algorithm Implementation} \label{Practic}
The efficiency and reliability of the proposed algorithm depend on the quality of the input data. Hence, certain hardware requirements must be satisfied to ensure its correct operation. First, the input measurements need to be unbiased and have a low noise level. This requires the use of optical current and voltage transformers since conventional transformers might have significant deviations of the secondary signals from the primary ones \cite{hrabluik_optical_2002}. Second, the proposed algorithm requires accurate estimation of time derivatives of the measured currents because it is based on analyzing differential equations of a power line model. To achieve this for an observation window of several milliseconds, a high sampling frequency in the range of dozens of kHz should be supported by a protection device. Third, since the algorithm relies on the exchange of measurements between the protection devices at two ends of the protected line, high-speed and high-bandwidth communication channels are required. Therefore, a fiber-optic channel or 5G communication is assumed to be utilized by the protection algorithm. It is worth noting that the aforementioned technologies are becoming increasingly available \cite{schweitzer_speed_2015}, which makes the proposed algorithm applicable in modern grids.


\section{Simulation Results} \label{Simulations}
The proposed algorithm was extensively tested for different types of generation, various levels of measurement noise, inaccuracies in assumed line parameters and different operating conditions: normal operation, internal and external faults.

\subsection{Simulation Setup} \label{Simulation_setup}
\begin{figure}[!t] 
	\centering
	\vspace{-0.03in}
	\includegraphics  [width=0.33\textwidth]{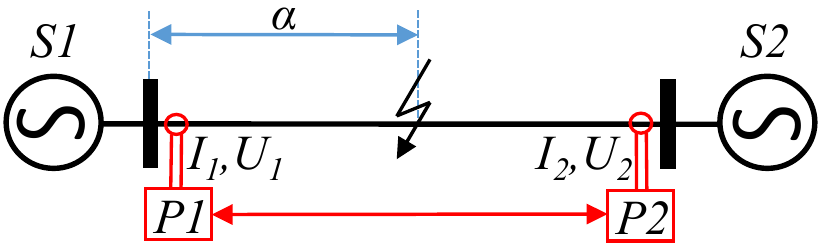}
	\vspace{-0.1in}
	\caption{Grid model for evaluation of proposed algorithm}
	\label{fig:sim_grid}
	\vspace{-0.1in}
\end{figure}

The proposed algorithm was evaluated on the grid representation shown in Fig.~\ref{fig:sim_grid}, with the protected line modeled as described in Section~\ref{Line_mod}. The solidly grounded supplying systems $S1$ and $S2$ were modeled as voltage sources with magnitudes $E_1$ and $E_2$, angle difference $\theta$ and coupled $RL$ branches. The protection system of a power line consists of two relays $P1$ and $P2$ that are connected by a communication channel and run the proposed algorithm. Since $P1$ and $P2$ are identical, only the operation of $P1$ was analyzed in detail. The grid model and the proposed algorithm were implemented in Simulink and MATLAB, respectively. Table~\ref{table:grid_data} shows the considered ranges of the parameters of the grid components. The parameters of the protected line include the line length $d$, the positive sequence resistance and inductance ${R}_L$ and ${L}_L$, respectively, and the ratio $K_{L}$ between the zero and positive sequence resistances and inductances. Similar parameters are defined for the models of the equivalent systems $S1$ and $S2$. Table~\ref{table:grid_data} also provides the considered grid voltage level $U_g$ and its power frequency $f$. A fault in the grid was modeled as shown in Fig.~\ref{fig:line_model}. The considered fault types and corresponding ranges of fault characteristics are presented in Table~\ref{table:fault_type}. Here, $T$ is the fault inception moment, $K_3$, $K_2$, $K_{2g}$ and $K_1$ denote three-phase, two-phase without ground, two-phase with ground and one-phase-to-ground fault types, respectively.    \par

\begin{table}[t!]
	\centering
	\caption{Parameters of test grid components}
	\vspace{-0.05in}
	\label{table:grid_data}
	\small
	\setlength{\tabcolsep}{2 pt}
	\begin{tabular}{|c|c|c|c|c|c|}
		\hline
		\multicolumn{4}{|c|}{Protected Line}                         & \multicolumn{2}{c|}{Grid} \\ \hline
		$L_L$, mH/km    & $R_L$, $\Omega$/km  & $K_L$, p.u. & $d$, km  & $U_g$, kV     & $f$, Hz   \\ \hline
		{[}1.3, 1.4{]} & {[}0.2, 0.42{]}  & 3  & {[}10, 80{]} & 36              & 50      \\ \hline
		\multicolumn{6}{|c|}{Systems $S1$ and $S2$}                                                      \\ \hline
		$L_S$, mH       & $R_S$, $\Omega$     & $K_S$, p.u.   & $E_1$, p.u.  &    $E_2$, p.u.  & $\theta$, $^\circ$                  \\ \hline
		{[}46, 250{]}  & {[}1.4, 19.4{]} & 1.5        & {[}0.9, 1.1{]}  & {[}0.9, 1.1{]}  & {[}-30, 30{]}    \\ \hline
	\end{tabular} \vspace{-0.15in}
\end{table}

\begin{table}[t!]
	\centering
	\caption{Considered fault types and ranges of fault characteristics}
	\vspace{-0.05in}
	\label{table:fault_type}
	\small
	\setlength{\tabcolsep}{6.4 pt}
	\begin{tabular}{|c|c|c|c|c|c|}
		\hline
		Type & $R_a$, $\Omega$   & $R_b$, $\Omega$   & $R_c$, $\Omega$   & $R_g$, $\Omega$ & $T$, ms     \\ \hline
		$K_3$   & \multicolumn{3}{c|}{$R_a=R_b=R_c \in $   {[}0, 100{]}}              & $\infty$   & {[}5, 25{]}       \\ \hline
		$K_2$   & \multicolumn{2}{c|}{$(R_a+R_b) \in $   {[}0, 100{]}} & $\infty$    & $\infty$ & {[}5, 25{]}        \\ \hline
		$K_{2g}$  & {[}0, 100{]}         & {[}0, 100{]}             & $\infty$          & {[}0, 100{]} & {[}5, 25{]} \\ \hline
		$K_1$   & {[}0, 100{]}             & $\infty$                      & $\infty$          & 0 & {[}5, 25{]}    \\ \hline
	\end{tabular} \vspace{-0.2in}
\end{table}

In the proposed algorithm, the sampling rate was set to 100~kHz and the derivatives of measured currents were obtained using subsets of $l=5$ adjacent samples. The upper bound on the elements of vector $x$ was set to $x^{\max}:=\left[\, \infty  \, , \, \infty \, , \, \infty \, , \, \infty \, , \, 1 \,\right]^T$. This bound was utilized in optimization problems \eqref{eq:opt_fault}, \eqref{eq:fault} and \eqref{eq:opt_mixed}, which were solved with Gurobi through its MATLAB interface \cite{gurobi}. The length of the observation window was set to 2~ms, which corresponds to 200 timestamps, and the number of considered mixture cases was set to $M=10$. The successive observation windows were assumed to be adjacent to each other, which minimizes the time between the fault inception and its detection. \par


\subsection{Normal Grid Operation Conditions} \label{Sim_Norm}
Initially, the security of the proposed algorithm was evaluated on one hundred scenarios corresponding to normal operating conditions of the grid. For each scenario, the parameters of grid components were chosen randomly from the ranges provided in Table~\ref{table:grid_data} and the grid model was simulated for 20~ms. Then, the measurements obtained from both line ends were fed to $P1$, where the proposed algorithm processed the measurement windows one by one. For all considered scenarios and all observation windows, the proposed algorithm correctly identified that the protected line was healthy for the entire simulation period.   

\subsection{External faults} \label{Sim_ext}
Next, a number of tests were carried out to assess the security of the proposed algorithm in case of external faults. For this, one hundred sets of parameter values for the grid components were selected randomly from the ranges given in Table~\ref{table:grid_data}. For each set, a large number of external fault instances were simulated. The diversity of these instances was ensured by simulating faults with all possible combinations of their characteristics within the limits defined in Table~\ref{table:fault_type} with the following step sizes: $\Delta R_a=\Delta R_b= \Delta R_c= \Delta R_g= 10 \Omega$ and $\Delta T = 0.5$ ms. Only external faults located at the buses incident to the protected line were examined as they are the most critical ones. These faults were assumed to be cleared by bus protection within 15 to 30 ms after their inception, which was chosen randomly for each simulation. The total simulation time was set to 70 ms. The measurements acquired after each simulation were analyzed by the proposed algorithm in the same way as described in the previous subsection. For all considered tests and observation windows, the proposed algorithm correctly identified that the protected line was healthy for the entire simulation time. 

\subsection{Internal faults} \label{Sim_int}
The dependability of the proposed algorithm for internal faults ($\alpha\! \in\![0,1]$) was assessed using one hundred sets of different values for the parameters of the grid components, which were selected randomly from Table~\ref{table:grid_data}. For each set, internal faults were simulated with all possible combinations of their characteristics within the limits defined in Table~\ref{table:fault_type} and the following step sizes: $\Delta R_a=\Delta R_b= \Delta R_c= \Delta R_g= 10 \Omega$, $\Delta \alpha=0.1$ p.u. and $\Delta T = 0.5$ ms. The total simulation time of the grid model was set to 32 ms. The measurements acquired after each simulation were analyzed by the proposed algorithm in the same way as described above. For presentation clarity, the obtained results are classified into three groups: 
\begin{itemize}
	\item group 1: the observation windows that contained samples corresponding only to the pre-fault conditions;
	\item group 2: the observation windows that contained the inception moment of an internal fault;
	\item group 3: the observation windows that contained samples corresponding only to after-fault conditions.
\end{itemize} 

\subsubsection{Group 1} \label{Sim_int_no} For all considered observation windows in this group, the proposed algorithm correctly identified that the protected line was healthy. 

\subsubsection{Group 2} \label{Sim_int_mix} The results for the second group are presented in Table~\ref{table:int_fault}. As can be seen, the proposed algorithm correctly detected an internal fault for all considered fault instances. This table also presents the maximum and mean estimation errors of the fault location and fault resistances for each fault type. Note that the mean errors of fault resistances were calculated by averaging the maximum errors of estimating $R_a$, $R_b$, $R_c$ and $R_g$ over all tests.

\begin{table}[t!]
	\centering
	\caption{Results for internal faults for base test cases (Section \ref{Sim_int})}
	\vspace{-0.05in}
	\label{table:int_fault}
	\small
	\setlength{\tabcolsep}{3.1 pt}
\begin{tabular}{|c|c|c|c|c|c|c|}
	\hline
	\multicolumn{1}{|l|}{\multirow{3}{*}{Group}} & \multirow{3}{*}{\begin{tabular}[c]{@{}c@{}}Fault \\ Type\end{tabular}} & Protection                                                                       & \multicolumn{4}{c|}{Fault Characteristics}                                                    \\ \cline{3-7} 
	\multicolumn{1}{|l|}{}                       &                                                                        & \multirow{2}{*}{\begin{tabular}[c]{@{}c@{}}\% Faults\\ Detected\end{tabular}} & \multicolumn{2}{c|}{Location Error, m}        & \multicolumn{2}{c|}{Resistance Error, $\Omega$}     \\ \cline{4-7} 
	\multicolumn{1}{|c|}{}                       &                                                                        &                                                                                  & max                   & mean                  & max                   & mean                  \\ \hline
	\multirow{4}{*}{2}                           & $K_3$ & 100         & 6.66     &  0.78    & 9.13            &  0.009         \\ \cline{2-7} 
	& $K_2$                                                                     & 100  &   48.71   &  0.38      &  1.98  &   0.005      \\ \cline{2-7} 
	& $K_{2g}$                                                                    & 100     & 30.99  &  0.65 & 0.41 &    0.031                \\ \cline{2-7} 
	& $K_1$   & 100   &   6.57  &  0.54 &   0.09  & 0.006       \\ \hline
	\multirow{4}{*}{3}                           & \multicolumn{1}{c|}{$K_3$}                                                  & \multicolumn{1}{c|}{100}                                                            & \multicolumn{1}{c|}{4.27} & \multicolumn{1}{c|}{0.59} & \multicolumn{1}{c|}{0.29} & \multicolumn{1}{c|}{0.002} \\ \cline{2-7} 
	& \multicolumn{1}{c|}{$K_2$}                                                  & \multicolumn{1}{c|}{100}                                                            & \multicolumn{1}{c|}{18.31} & \multicolumn{1}{c|}{0.23} & \multicolumn{1}{c|}{0.01} & \multicolumn{1}{c|}{$<$0.001} \\ \cline{2-7} 
	& \multicolumn{1}{c|}{$K_{2g}$}                                                  & \multicolumn{1}{c|}{100}                                                            & \multicolumn{1}{c|}{29.24} & \multicolumn{1}{c|}{0.42} & \multicolumn{1}{c|}{0.08} & \multicolumn{1}{c|}{0.006} \\ \cline{2-7} 
	& \multicolumn{1}{c|}{$K_1$}                                                  & \multicolumn{1}{c|}{100}                                                            & \multicolumn{1}{c|}{2.72} & \multicolumn{1}{c|}{0.29} & \multicolumn{1}{c|}{0.01} & \multicolumn{1}{c|}{$<$0.001} \\ \hline
\end{tabular} \vspace{-0.2in}
\end{table}

\subsubsection{Group 3} \label{Sim_int_fault} The results for the last group are also presented in Table~\ref{table:int_fault}, with the same criteria for assessing the algorithm's performance as for the second group. The algorithm correctly identified the state of the line as faulted for all considered faults. The maximum and mean estimation errors of fault location and resistances were significantly reduced for this group compared to the errors obtained for the second group. The reason for this is that the observation windows in group 3 contained more samples corresponding to after-fault conditions, which increased the estimation accuracy of the fault characteristics. Therefore, in practice the fault characteristics should be estimated using windows in this group. \par

The results obtained in Sections~\ref{Sim_Norm} to \ref{Sim_int} demonstrate that the proposed algorithm can perform reliable line protection and reasonably accurate estimation of fault characteristics. The simulation results also indicate that the proposed algorithm correctly estimated the fault inception interval for all simulated instances of internal faults.

It is worth mentioning that successive observation windows can be made adjacent to each other only if the combined time of data collection and algorithm execution is below the window's length of 2 ms. The data collection time comprises the delays of communication channel propagation and communication equipment. For the highest considered line length, it does not exceed 1 ms \cite{costa_two-terminal_2017}. Hence, the proposed algorithm should be executed in under 1 ms. While the MATLAB implementation resulted in a slightly higher runtime, it could be significantly reduced by using a lower level programming language and dedicated hardware of a microprocessor relay.

\subsection{Impact of converter-based generation sources} \label{Sim_conv}
Converter-based generators may introduce a number of challenges for conventional protection algorithms \cite{norshahrani_progress_2017}. Therefore, it is important to test the proposed algorithm on a grid model with this type of sources. Hence, source $S2$ in Fig.~\ref{fig:sim_grid} was substituted by a model of a wind farm consisting of five identical Type 4 wind turbines. The parameters of grid components were set as given in Table~\ref{table:Setup_conv}, where $P$ is the nominal power, $H$ is the inertia constant, $R_s$ is the stator resistance and $X_l$ is the leakage reactance. The values of d- and q-axis reactances are also reported in the table. All parameters for which no units are provided are given in per unit.
 
\begin{table}[t!]
	\centering
	\caption{Setup for test cases with converter-based generation (Section~\ref{Sim_conv})}
	\vspace{-0.05in}
	\label{table:Setup_conv}
	\small
	\setlength{\tabcolsep}{1.6 pt}
	\begin{tabular}{|c|c|c|c|c|c|c|c|c|}
		\hline
		\multicolumn{4}{|c|}{Protected Line}                 & \multicolumn{5}{c|}{$S1$}          \\ \hline
		$L_L$, H/km    & $R_L$, $\Omega$/km  & $K_L$ & $d$, km  &	
		$L_S$, H       & $R_S$, $\Omega$     & $K_S$   & $E_1$  &  $\theta_1$, $^\circ$ \\ \hline
		0.0013           & 0.2             & 3          & 40    & 0.046            & 1.4 & 1.5   & 1   & 0  \\ \hline
		\multicolumn{9}{|c|}{$S2$ (data for one wind turbine)}                                             \\ \hline
		$P$, MW & $X_d$  & $X'_d$ & $X''_d$ & $X_q$  & $X''_q$ & $X_l$ & $R_s$ & $H$, s \\ \hline
		2 & 1.305 & 0.296  & 0.252  & 0.474 & 0.243 & 0.18   & 0.006   & 0.62  \\ \hline
	\end{tabular} \vspace{-0.1in}
\end{table}

To assess the security of the proposed algorithm for external faults and normal operation of the grid, a number of external fault instances were simulated and analyzed as described in Section~\ref{Sim_ext}. For all considered tests and observation windows, the proposed algorithm correctly identified that the protected line was healthy for the entire simulation time.  \par

To assess the dependability of the proposed algorithm, a number of internal faults were simulated and analyzed as described in Section~\ref{Sim_int}. For the observation windows that did not have samples corresponding to fault conditions (group 1), the proposed algorithm correctly identified the protected line as healthy. For the rest of observation windows (groups 2 and 3), the results are presented in Table~\ref{table:int_fault_conv}. The proposed algorithm correctly detected the internal fault for all considered fault instances. The maximum and mean estimation errors of fault location and resistances had similar values and followed the same trend as in Section~\ref{Sim_int}. Note that the proposed algorithm also correctly estimated the fault inception interval for all simulated instances of internal faults.

\begin{table}[t!]
	\centering
	\caption{Results for internal faults for test cases with converter-based generation (Section~\ref{Sim_conv})}
	\vspace{-0.05in}
	\label{table:int_fault_conv}
	\small
	\setlength{\tabcolsep}{3.1 pt}
	\begin{tabular}{|c|c|c|c|c|c|c|}
		\hline
		\multicolumn{1}{|l|}{\multirow{3}{*}{Group}} & \multirow{3}{*}{\begin{tabular}[c]{@{}c@{}}Fault \\ Type\end{tabular}} & Protection                                                                       & \multicolumn{4}{c|}{Fault Characteristics}                                                    \\ \cline{3-7} 
		\multicolumn{1}{|l|}{}                       &                                                                        & \multirow{2}{*}{\begin{tabular}[c]{@{}c@{}}\% Faults\\ Detected\end{tabular}} & \multicolumn{2}{c|}{Location Error, m}        & \multicolumn{2}{c|}{Resistance Error, $\Omega$}     \\ \cline{4-7} 
		\multicolumn{1}{|c|}{}                       &                                                                        &                                                                                  & max                   & mean                  & max                   & mean                  \\ \hline
		\multirow{4}{*}{2}                           & $K_3$ & 100         &  30.83 &   5.79 & 1.45 &  0.036         \\ \cline{2-7} 
		& $K_2$   & 100   & 82.79  & 7.15  & 2.50  &  0.055  \\ \cline{2-7} 
		& $K_{2g}$                                                                    & 100 &   39.77 &  6.57  & 0.87    &  0.058   \\ \cline{2-7} 
		& $K_1$                                                                     & 100 &  21.43      &  6.69  & 0.78    & 0.021      \\ \hline
		\multirow{4}{*}{3}                           & \multicolumn{1}{c|}{$K_3$}                                                  & \multicolumn{1}{c|}{100}                                                            & \multicolumn{1}{c|}{17.16} & \multicolumn{1}{c|}{4.45} & \multicolumn{1}{c|}{0.94} & \multicolumn{1}{c|}{0.024} \\ \cline{2-7} 
		& \multicolumn{1}{c|}{$K_2$}                                                  & \multicolumn{1}{c|}{100}                                                            & \multicolumn{1}{c|}{40.84} & \multicolumn{1}{c|}{5.01} & \multicolumn{1}{c|}{1.72} & \multicolumn{1}{c|}{0.053} \\ \cline{2-7} 
		& \multicolumn{1}{c|}{$K_{2g}$}                                                  & \multicolumn{1}{c|}{100}                                                            & \multicolumn{1}{c|}{27.61} & \multicolumn{1}{c|}{4.83} & \multicolumn{1}{c|}{0.10} & \multicolumn{1}{c|}{0.023} \\ \cline{2-7} 
		& \multicolumn{1}{c|}{$K_1$}                                                  & \multicolumn{1}{c|}{100}                                                            & \multicolumn{1}{c|}{14.55} & \multicolumn{1}{c|}{5.69} & \multicolumn{1}{c|}{0.08} & \multicolumn{1}{c|}{0.011} \\ \hline
	\end{tabular}\vspace{-0.15in}
\end{table}

\subsection{Impact of measurement noise} \label{Sim_noise}
\begin{figure}[!t]   
	\centering
	\subfloat[Measure of security]{\includegraphics[width=1.65in]{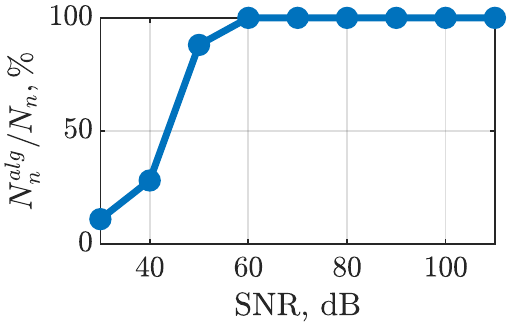}
		\label{fig_Sec}}
	\hfil
	\subfloat[Measure of dependability]{\includegraphics[width=1.65in]{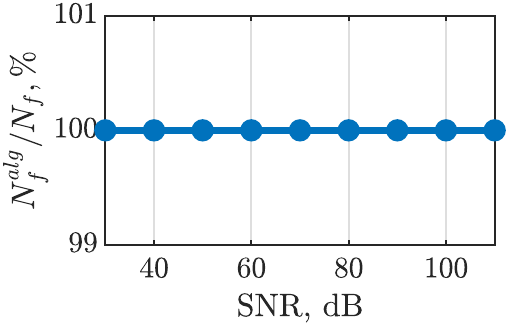}
		\label{fig_Dep}}
	\hfil
	\subfloat[Location error]{\includegraphics[width=1.65in]{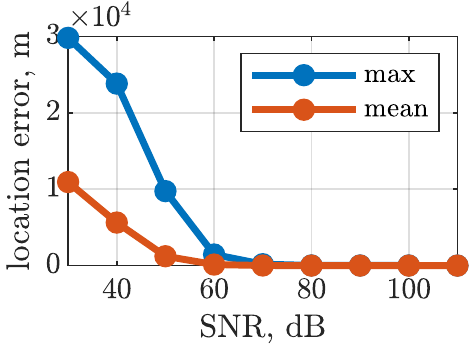}
		\vspace{-0.05in}
		\label{fig_Al}}
	\hfil
	\subfloat[Resistance error]{\includegraphics[width=1.65in]{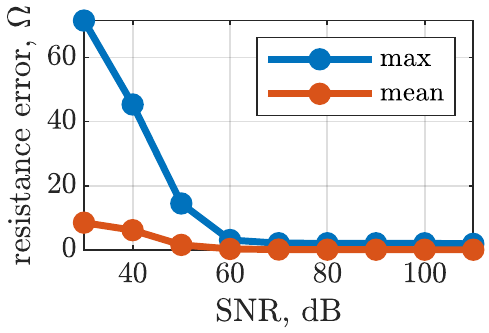}
		\label{fig_Rf}}
	\caption{Impact of measurement noise on algorithm performance}
	\label{fig_noise}
	\vspace{-0.2in}
\end{figure}

Despite the high accuracy of optical instrumental transformers and advanced filtering methods utilized in modern protection relays, some noise can still be present in signals processed by protection algorithms. Therefore, the proposed algorithm was tested using scenarios described in Sections~\ref{Sim_Norm} to \ref{Sim_int} and adding Gaussian noise to all measurements. Normal grid operation, external and internal faults were simulated for the same one hundred sets of the values of the grid component parameters, which were selected randomly from the ranges in Table~\ref{table:grid_data}. Figure \ref{fig_noise} presents the assessment of dependability and security of the proposed algorithm and estimation errors of fault location and resistances for different signal-to-noise ratios (SNRs) from 30 to 110 dB. Here, $N_f$ is the number of observation windows with samples corresponding to internal fault conditions, $N_f^{alg}$ is the number of times the proposed algorithm correctly identified an internal fault when analyzing these windows. Similarly, $N_n$ is the number of windows with samples corresponding only to normal grid conditions and/or external faults and $N_n^{alg}$ is the number of times the algorithm correctly identified the protected line state as healthy when analyzing these windows. The estimation errors in Fig.~\ref{fig_Al} and Fig.~\ref{fig_Rf} are shown for observation windows of group 3 since these windows lead to a more reliable evaluation of fault characteristics as described above. Figure \ref{fig_noise} shows the results only for $K_3$ faults; the other fault types led to similar results. \par

The results indicate that the dependability of the proposed algorithm was not affected by the considered increase in the noise level, whereas security was adequate only for SNR~$\geq$~60~dB. The fault location and estimation of fault resistances achieved reasonable accuracy approximately for the same range of SNRs. Therefore, the proposed algorithm was able to reliably perform all its functions for SNR~$\geq$~60~dB, which covers a realistic range of noise levels induced by modern optical instrumental transformers \cite{silva_optical_2012}. Note that the estimation error of the fault inception interval followed a similar trend to the one exhibited by the measure of security.     

\subsection{Impact of inaccuracies in line parameters} \label{Sim_line_inacc}
\begin{figure}[!t]   
	\vspace{-0.15in}
	\centering
	\subfloat[Max]{\includegraphics[width=1.65in]{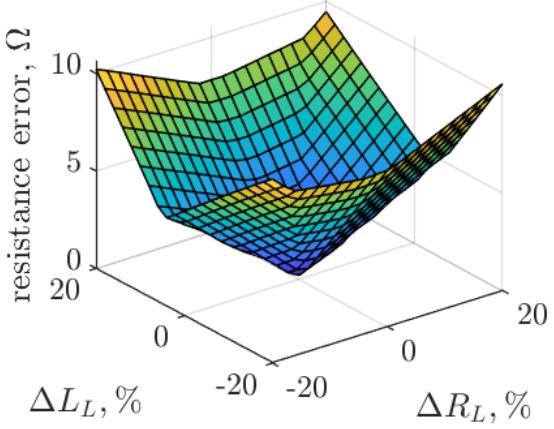}
		\label{fig_Rf_max}}
	\hfil
	\subfloat[Mean]{\includegraphics[width=1.65in]{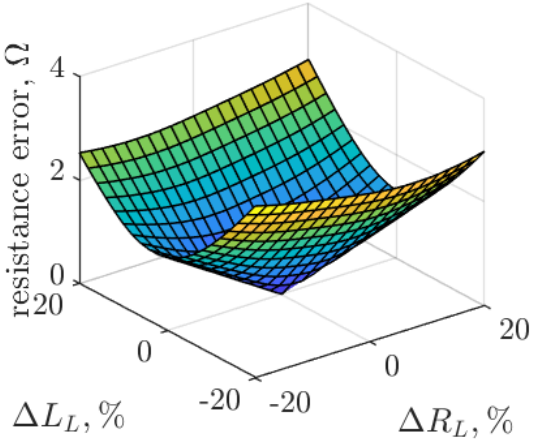}
		\label{fig_Rf_mean}}
	\hfil
	\subfloat[Max]{\includegraphics[width=1.65in]{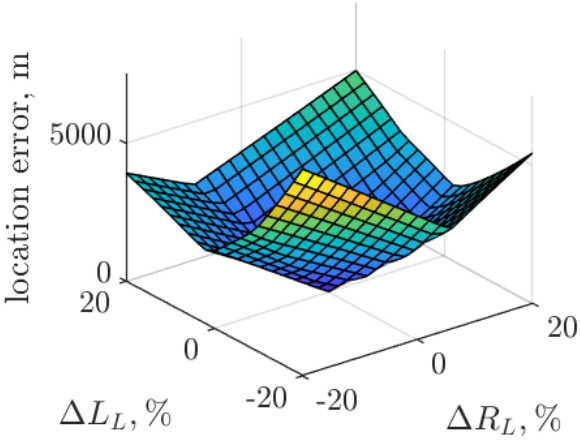}
		\vspace{-0.05in}
		\label{fig_Al_max}}
	\hfil
	\subfloat[Mean]{\includegraphics[width=1.65in]{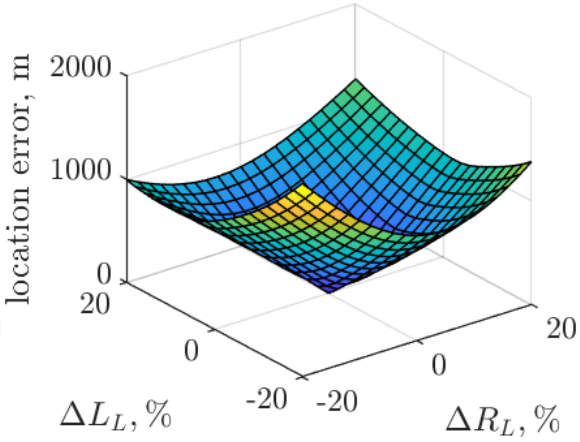}
		\label{fig_Al_mean}}
	\caption{Impact of uncertainty in line parameters on algorithm performance}
	\label{fig_line_inaccuracy}
	\vspace{-0.15in}
\end{figure}

Finally, the robustness of the proposed algorithm to the inaccuracies in the input line parameters $L_L$ and $R_L$ was evaluated. This was done by simulating the same scenarios as described in Section~\ref{Sim_noise} but for fixed line parameters provided in Table~\ref{table:Setup_conv} and infinite SNR. The input values of $L_L$ and $R_L$ fed to the algorithm were changed by up to $\pm 20 \%$ compared to the values utilized for the grid simulation. For all considered scenarios the algorithm was secure and dependable. Figure \ref{fig_line_inaccuracy} presents the estimation errors of fault location and resistances for different deviations of line parameters utilized in the algorithm from the parameters employed for the simulations. The fault location and estimation of fault resistances achieved reasonable accuracy for $L_L$ and $R_L$ in the range of $\pm 10 \%$, which is larger than the accuracy of available methods for line parameters' estimation \cite{costa_estimation_2015,liao_online_2009}. Note that the proposed algorithm also correctly estimated the fault inception interval for all simulated internal fault instances.

\section{Conclusion} \label{Conclusion}
This paper presents a settingless optimization-based unit protection algorithm for medium-voltage power lines. The fact that no set-points need to be defined simplifies the algorithm's application and reduces the risk of human errors. The algorithm combines protection, fault location and fault type identification functionalities, which allows for the unification of these closely related functions. The main idea of the proposed algorithm is to identify which model of a protected line, healthy or with an internal fault, best fits the input measurements. For that purpose, a number of convex optimization problems are solved, which allows to reliably detect an internal fault and estimate its characteristics that are most consistent with the data. \par

The proposed algorithm was extensively tested on a high number of scenarios.
They included different levels of measurement noise, uncertainty of line parameters, and presence of converter-based generation in the grid. The results demonstrate that the algorithm was secure and dependable for all considered scenarios with realistic noise levels even without utilization of filtering methods. The algorithm also achieved accurate estimation of characteristics of internal faults for realistic levels of noise and errors in the assumed line parameters, even for high-impedance faults. \par

An important direction of future work is to make the proposed algorithm applicable for high voltage transmission lines, which require more detailed modeling of the lines.






\bibliographystyle{IEEE}
\bibliography{MB}

\end{document}